\documentclass[twocolumn,showpacs,aps,groupedaddress,prd,eqsecnum]{revtex4}
\usepackage{amsfonts,latexsym,eucal}
\usepackage[dvips]{graphicx}

\def\beq{\begin{equation}}
\def\eeq{\end{equation}}
\def\bea{\begin{eqnarray}}
\def\eea{\end{eqnarray}}

\def\pa{\partial}
\def\ra{\rightarrow}

\def\tt{\tilde{t}}
\def\tr{\tilde{r}}
\def\tp{\tilde{\phi}}
\def\tpp{{\tp_{~}}'}
\def\tpps{{\tp_{~}}^{\prime2}}
\def\tv{\tilde{V}}

\def\to{\tilde{\omega}}
\def\bp{\mbox{\boldmath$\phi$}}
\def\tE{\tilde{E}}

\usepackage{epsfig}

\begin{document}

\thispagestyle{empty}

\title{What are universal features of gravitating Q-balls?}
\author{Takashi Tamaki}
\email{tamaki@ge.ce.nihon-u.ac.jp}
\affiliation{Department of Physics, General Education, College of Engineering, 
Nihon University, Tokusada, Tamura, Koriyama, Fukushima 963-8642, Japan}
\author{Nobuyuki Sakai}
\email{nsakai@e.yamagata-u.ac.jp}
\affiliation{Department of Education, Yamagata University, Yamagata 990-8560, Japan}
\date{\today}

\begin{abstract}
We investigate how gravity affects Q-balls by exemplifying the case of
the Affleck-Dine potential $V(\phi):=m^4 \ln (1+\frac{\phi^2}{m^2})$. 
Surprisingly, stable Q-balls with arbitrarily small charge exist, 
no matter how weak gravity is, contrary to the case of flat spacetime. 
We also show analytically that this feature holds true for general models as long as 
the leading order term of the potential is a positive mass term in its Maclaurin series.
\end{abstract}

\maketitle

\section{Introduction}
Q-balls \cite{Col85}, a kind of nontopological solitons \cite{LP92}, appear in a large family of field 
theories with global U(1) (or more) symmetry. It has been argued that 
Q-balls with the Affleck-Dine (AD) potential could play important roles in cosmology \cite{AD}.
For example, Q-balls can be produced efficiently and could be responsible for baryon 
asymmetry \cite{SUSY} and dark matter \cite{SUSY-DM}. 
Therefore, their stability is an important subject to be studied.
Because Q-balls are typically supposed to be microscopic objects, their self-gravity is usually ignored; and accordingly, stability of Q-balls with various potentials has been intensively studied in flat spacetime \cite{stability,PCS01,SakaiSasaki,Copeland}. 

However, if we contemplate the results on boson stars~\cite{boson-review}, we notice the possibility that 
self-gravity can be important even if it is very weak.   
For example, in the case of the potential $V_{\rm mini}=m^2\phi^2/2$, equilibrium solutions, called (mini-)boson stars, exist due to self-gravity, though no equilibrium solution exists without gravity.
An important point is that there is no minimum charge for (mini-)boson stars, i.e., they can exist 
even if their self-gravity is very weak. This example suggests the importance of the unified picture of 
Q-balls and boson stars, which are different from each other solely in model parameters.

From this motivation, we have investigated the following three models.
The first one is~\cite{TamakiSakai} 
\beq\label{v3}
V_{3}(\phi):={m^2\over2}\phi^2-\mu\phi^3+\lambda\phi^4 ~~~
{\rm with} ~~~ m^2,~\lambda>0\ , 
\eeq
which describe Q-balls ($\mu >0$) and boson stars ($\mu =0$) comprehensively.
The second one is~\cite{TamakiSakai2} 
\beq\label{V4}
V_4(\phi):={m^2\over2}\phi^2-\lambda\phi^4+\frac{\phi^6}{M^2} 
~~~{\rm with} ~~~ m^2,~\lambda,~M>0.
\eeq
Interestingly, gravitating Q-balls (boson stars) in these two models have a common property that 
stable Q-balls with arbitrarily small charge exist no matter how weak self-gravity is,
while Q-ball properties in flat spacetime fairly depend on the potentials.  

Although these examples reveal gravitational effects on Q-balls, both models are described by polynomial in 
$\phi$. 
From the theoretical point of view, however, models with logarithmic terms are more natural.
Specifically, in the AD mechanism noted above, there are two types of potentials:
gravity-mediation type and gauge-mediation type. 
The former type is described by the potential,
\beq\label{gravity}
V_{\rm grav.}(\phi):=\frac{m^2}{2}\phi^2\left[
1+K\ln \left(\frac{\phi}{M}\right)^2
\right]~~~
{\rm with} ~~~ m^2,~M>0\ .
\eeq
In flat spacetime equilibrium solutions for $K\geq 0$ are nonexistent while those for $K<0$ are 
existent. 
If we take self-gravity into account, stable Q-balls exist even for $K=0$
since the potential coincides with $V_{\rm mini}$. 
In our previous paper~\cite{TamakiSakai3}, 
we have shown that gravitating ``Q-balls" exist, which are surrounded by Q-matter, 
even for $K>0$. Unfortunately, since we cannot define the Q-ball charge in this case, 
it is difficult to say about the common property noted above and 
their stability has not been explored yet.

In the present paper we extend our analysis to the gauge-mediation type, 
\beq\label{gauge}
V_{\rm gauge}(\phi):=m^4 \ln\left(1+\frac{\phi^2}{m^2}\right)~~~{\rm with} ~~~ m>0\ .
\eeq
We shall show that gravitating Q-balls with (\ref{gauge}) 
have a common property with $V_{\rm mini}$, $V_{3}$ and $V_{4}$, and discuss the reason. 

This paper is organized as follows.
In Sec. II, we derive equilibrium field equations. 
In Sec. III, we show numerical results of equilibrium Q-balls. 
In Sec. IV, we explain why self-gravity gives a rather model independent feature even if 
it is very weak. In Sec. V, we devote to concluding remarks.

\section{Analysis method of equilibrium Q-balls}

\subsection{Equilibrium field equations}

We begin with the action
\beq\label{Sg}
{\cal S}=\int d^4x\sqrt{-g}\left\{ \frac{{\cal R}}{16\pi G}-\frac12g^{\mu\nu}\pa_{\mu}\bp\cdot\pa_{\nu}\bp
-V(\phi)\right\}, 
\eeq
where $\bp=(\phi_1,~\phi_2)$ is an SO(2)-symmetric scalar field and 
$\phi:=\sqrt{\bp\cdot\bp}=\sqrt{\phi_1^2+\phi_2^2}$.
We assume a spherically symmetric and static spacetime, 
\beq\label{metric1}
ds^2=-\alpha^2(r)dt^2+A^2(r)dr^2+r^2(d\theta^2+\sin^2\theta d\varphi^2).
\eeq

For the scalar field, we assume that it has a spherically symmetric and stationary form, 
\beq\label{phase}
(\phi_1,\phi_2)=\phi(r)(\cos\omega t,\sin\omega t).
\eeq
Then the field equations become
\bea\label{Gtt}
-{r A^3\over2}G^t_t&:=&A'+{A\over2r}(A^2-1) \nonumber \\
&=&{4\pi G}r A^3\left({{\phi'}^2\over2A^2}
+{\omega^2\phi^2\over2\alpha^2}+V\right),
\\\label{Grr}
{r\alpha\over2}G_{rr}&:=&\alpha'+{\alpha\over2r}(1-A^2) \nonumber \\
&=&{4\pi G}r\alpha A^2
\left({{\phi'}^2\over2A^2}+{\omega^2\phi^2\over2\alpha^2}-V\right),
\\\label{Box}
{A^2\phi\over\phi_1}\Box\phi_1&:=&
\phi''+\left(\frac2r+{\alpha'\over\alpha}-{A'\over A}\right)\phi'
+\left({\omega A\over\alpha}\right)^2\phi \nonumber \\
&=&A^2{dV\over d\phi},
\eea
where $':= d/dr$. 
To obtain Q-ball solutions in curved spacetime, we should solve 
(\ref{Gtt})-(\ref{Box}) with boundary conditions, 
\bea
&& A(0)=A(\infty)=\alpha(\infty)=1,\nonumber \\ 
&& A'(0)=\alpha'(0)=\phi'(0)=\phi(\infty)=0.
\label{bcg}
\eea
We also restrict our solutions to monotonically decreasing $\phi (r)$. 
Because of the symmetry, there is a conserved charge called Q-ball charge,
\bea\label{Q}
Q&:= &\int d^3x\sqrt{-g}g^{0\nu}(\phi_1\pa_\nu\phi_2-\phi_2\pa_\nu\phi_1)=\omega I,
\nonumber  \\
&&{\rm where}~~~
I:=4\pi\int{A r^2\phi^2\over\alpha}dr.
\eea

We suppose $V_{\rm gauge}$ Model (\ref{gauge}).
Rescaling the quantities as
\bea
&&\tp\equiv\frac{\phi}{m},~~\tv_{\rm gauge}\equiv\frac{V_{\rm gauge}}{m^4}= \ln (1+\tp^2 ),
\nonumber  \\
&&\to\equiv\frac{\omega}{m},~~\kappa=Gm^2,~~\tt\equiv mt,~~ \tr\equiv mr,~~
\label{rescale-gauge}
\eea
the field equations (\ref{Gtt})-(\ref{Box}) are rewritten as
\beq\label{rsfe1}
A'+{A\over2\tr}(A^2-1)
=4\pi\kappa\tr A^3\left({\tpps\over2A^2}+{\to^2\tp^2\over2\alpha^2}+\tv_{\rm gauge}\right),
\eeq\beq\label{rsfe2}
\alpha'+{\alpha\over2\tr}(1-A^2)
=4\pi\kappa\tr\alpha A^2\left({\tpps\over2A^2}+{\to^2\tp^2\over2\alpha^2}-\tv_{\rm gauge}\right),
\eeq\beq\label{rsfe3}
\tp^{\prime\prime}+\left(\frac2{\tr}+{\alpha'\over\alpha}-{A'\over A}\right)\tpp
+\left({\to A\over\alpha}\right)^2\tp=A^2{d\tv_{\rm gauge}\over d\tp}.
\eeq

\subsection{Equilibrium solutions in flat spacetime}

In preparation for discussing gravitating Q-balls, we review their 
equilibrium solutions in flat spacetime ($\kappa=0$).
The scalar field equation (\ref{rsfe3}) reduces to
\beq\label{rsfeflat}
\tp^{\prime\prime}=-\frac{2}{\tilde{r}}\tpp-\tilde{\omega}^2\tilde{\phi}
+{d\tilde{V}_{\rm gauge}\over d\tilde{\phi}}\,.
\eeq
This is equivalent to the field equation for a single static scalar 
field with the potential $V_{\omega}:=\tilde{V}_{\rm gauge}-\tilde{\omega}^2\tilde{\phi}^2/2$.
Equilibrium solutions satisfying boundary conditions (\ref{bcg}) exist if 
\beq\label{excon}
{\rm min}(V_{\omega})<\tilde{V}_{\rm gauge}(0)~~{\rm and}~~
{d^2V_{\omega}\over d\tilde{\phi}^2}(0)>0.
\eeq
We obtain 
\bea
\frac{dV_{\omega} }{d\tp}=\frac{2\tp}{1+\tp^2}-\to^2\tp\ , \label{origin1} \\
\frac{d^2 V_{\omega} }{d\tp^2}=-\frac{4\tp^2}{(1+\tp^2)^2}+\frac{2}{1+\tp^2}-\to^2 \ . 
\label{origin2}
\eea
The first condition in (\ref{excon}) is trivially satisfied since 
$V_{\omega}$ is unbounded from below. If we introduce $\epsilon^2 :=2-\to^2$, 
the second condition in (\ref{excon}) leads to
\beq\label{condition1}
\epsilon^2 >0\ . 
\eeq
The two limits $\epsilon^2 \ra 2$ and $\epsilon^2 \ra 0$ correspond to the thin-wall limit 
and the thick-wall limit, respectively.

If one regards the radius $r$ as ``time" and the scalar amplitude $\phi(r)$ as 
``the position of a particle", one can understand Q-ball solutions in words of Newtonian 
mechanics.
Equation (\ref{rsfeflat}) describes a one-dimensional motion of a particle under 
the conserved force due to the potential $-V_{\omega}(\phi)$ and the ``time"-dependent 
friction $-(2/r)d\phi/dr$. 

To discuss gravitational effects later, it is useful to estimate the central value 
$\phi_{0}:=\tp (0)$ in flat spacetime. 
Because $V_{\omega}\approx0$ at $r=0$, its order of magnitude 
is estimated as a solution of $V_{\omega}=0$ ($\tp (0)\neq 0$). 
For $V_{\rm gauge}$ with the thick-wall condition $\epsilon^2 \ll 1$, we obtain 
\beq\label{phi0-flat}
\phi_{0}^2 \left(1-\frac{\to^2}{2}\right)-\frac{\phi_{0}^4}{2}\simeq 0, 
\eeq
where we have used Maclaurin expansion of $\ln (1+\tp^2)$ and 
neglected higher order terms $O(\phi_{0}^{5})$. 
Then, we obtain 
\beq\label{phi0-flat2}
\phi_{0}\simeq \epsilon.
\eeq
which verifies our assumption that higher order of $\phi_0$ is negligible.

\section{Gravitating Q-balls}

\begin{figure}[htbp]
\psfig{file=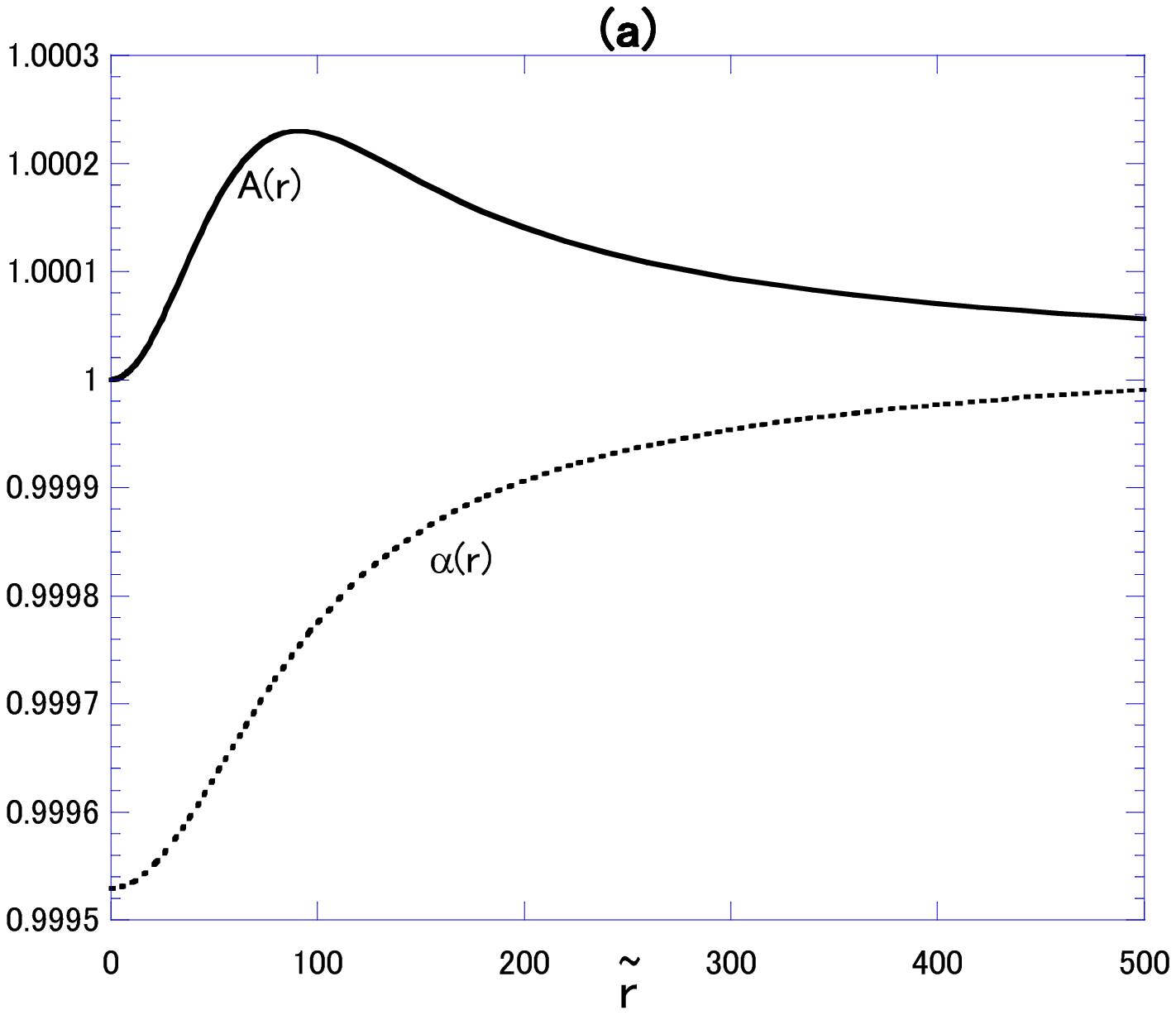,width=3in} \\
\psfig{file=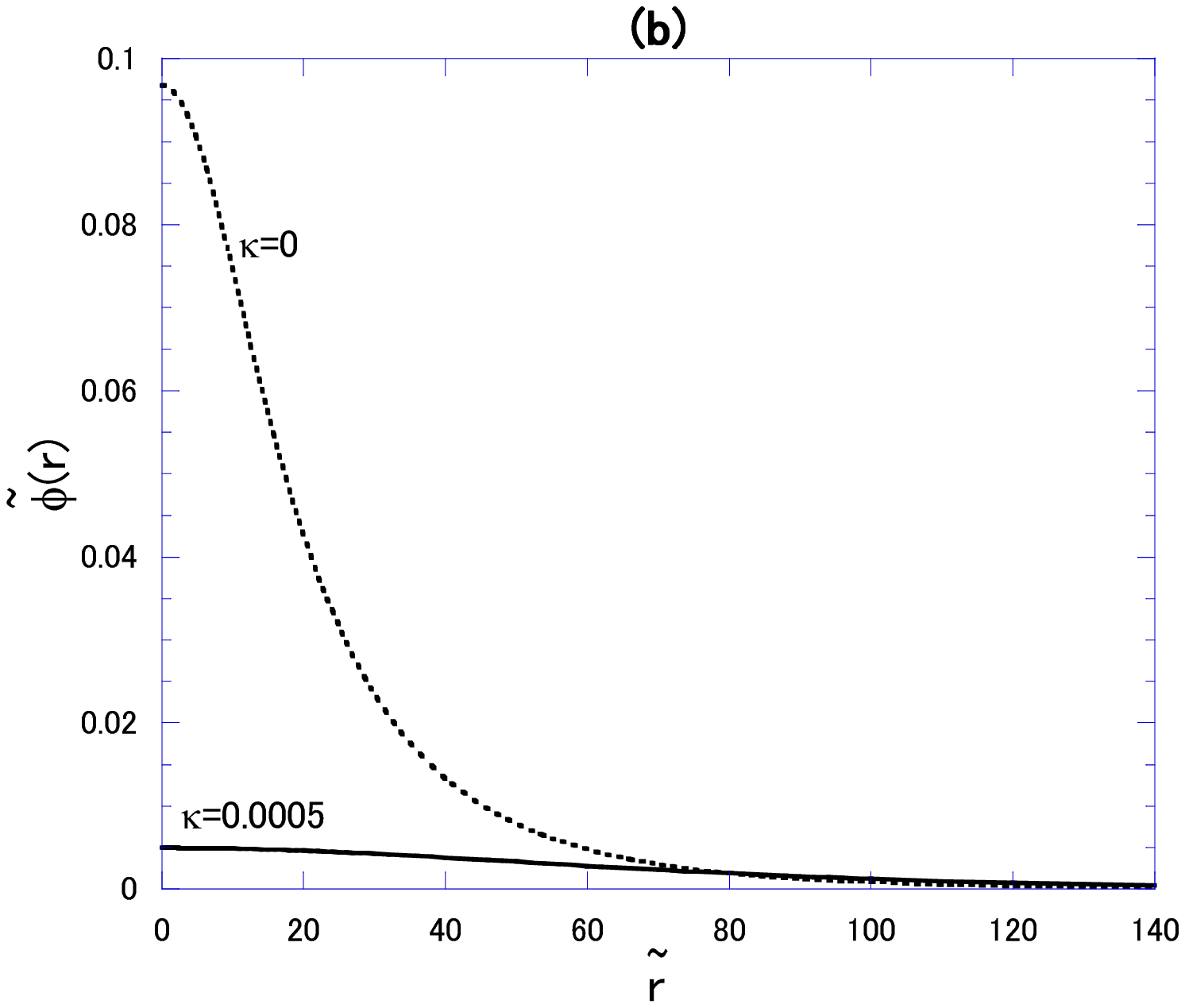,width=3in}
\caption{For a thick-wall Q-ball for $\to^2 \simeq 1.999$, we show 
(a) behavior of the metric functions for a gravitating one for $\kappa =0.0005$ and 
(b) comparison of $\tp (r)$ with that for the flat case $\kappa =0$. 
\label{thick-behavior} }
\end{figure}

The potential picture described above is effective in arguing equilibrium solutions also
in curved spacetime.
In this case, $V_{\omega}$ and $\epsilon^2$ should be redefined by 
\bea\label{thick2}
&&V_{\omega}:=\tv_{\rm gauge}-{\to^2\over2\alpha^2}\tp^2,~~~
\epsilon^{2} :=2-\frac{\to^2}{\alpha^2} . \label{origin4} 
\eea

``The potential of a particle", $-V_{\omega}$, is now  ``time"-dependent, 
which sometimes plays an important role, as we see below. 

First, we show the result on a thick-wall Q-ball with $\to^2 \simeq 1.999$ in Fig.~\ref{thick-behavior}.
We put  $\kappa =0.0005$ for a gravitating Q-ball.
The metric functions $\alpha(\tr),~A(\tr)$ and the field amplitude $\tp(\tr)$ are shown in (a) and (b), respectively.
Because $\alpha(\tr)$ and $A(\tr)$ are close to one, one may think that gravity acts as small perturbations.
However, looking at $\tp(\tr)$, we find that gravity changes its shape drastically.
Near the origin, we find that the scalar field for the gravitating case 
takes much smaller value than that for the flat case. 

\begin{figure}[htbp]
\psfig{file=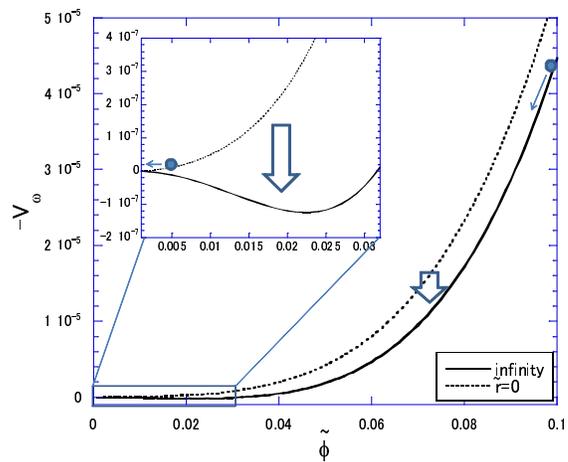,width=3in} 
\caption{$-V_{\omega}$ for a thick-wall Q-ball $\to^2 \simeq 1.999$. 
For the flat case $\kappa =0$, $-V_{\omega}$ is fixed as shown by a 
solid line. On the other hand, for the gravitating case $\kappa =0.0005$, $-V_{\omega}$ changes 
from a dotted line to a solid line as ``time" $\tr$ goes. 
As a result, the scalar field should start from $\tp \simeq 9.68\times 10^{-2}$ and 
$\tp \simeq 4.96\times 10^{-3}$ to satisfy $\tp (\infty )=0$ for the flat case and 
the gravitating case, respectively as shown in this figure. 
\label{Vw1999} }
\end{figure}

We explain the reason for this result by using the potential $-V_{\omega}$, as shown in Fig.~\ref{Vw1999}.
For the flat case $\kappa =0$, $-V_{\omega}$ is fixed as shown by a solid line.
As a result, the scalar field with a relatively large value ($\tp \simeq 9.68\times 10^{-2}$) at the initial time $\tr =0$ rolls down and climb up the potential and finally reaches $\tp =0$ at the time $\tr \ra \infty$. 
On the other hand, for the gravitating case $\kappa =0.0005$, $-V_{\omega}$ changes 
from a dotted line to a solid line as ``time" $\tr$ goes. 
An important point is that the sign of $-\frac{dV_{\omega}}{d\tp}$ changes near the origin. 
Therefore, the scalar field should take a sufficient small value ($\tp \simeq 4.96\times 10^{-3}$) to satisfy $\tp (\infty )=0$, as shown in Fig.~\ref{Vw1999}. 

\begin{figure}[htbp]
\psfig{file=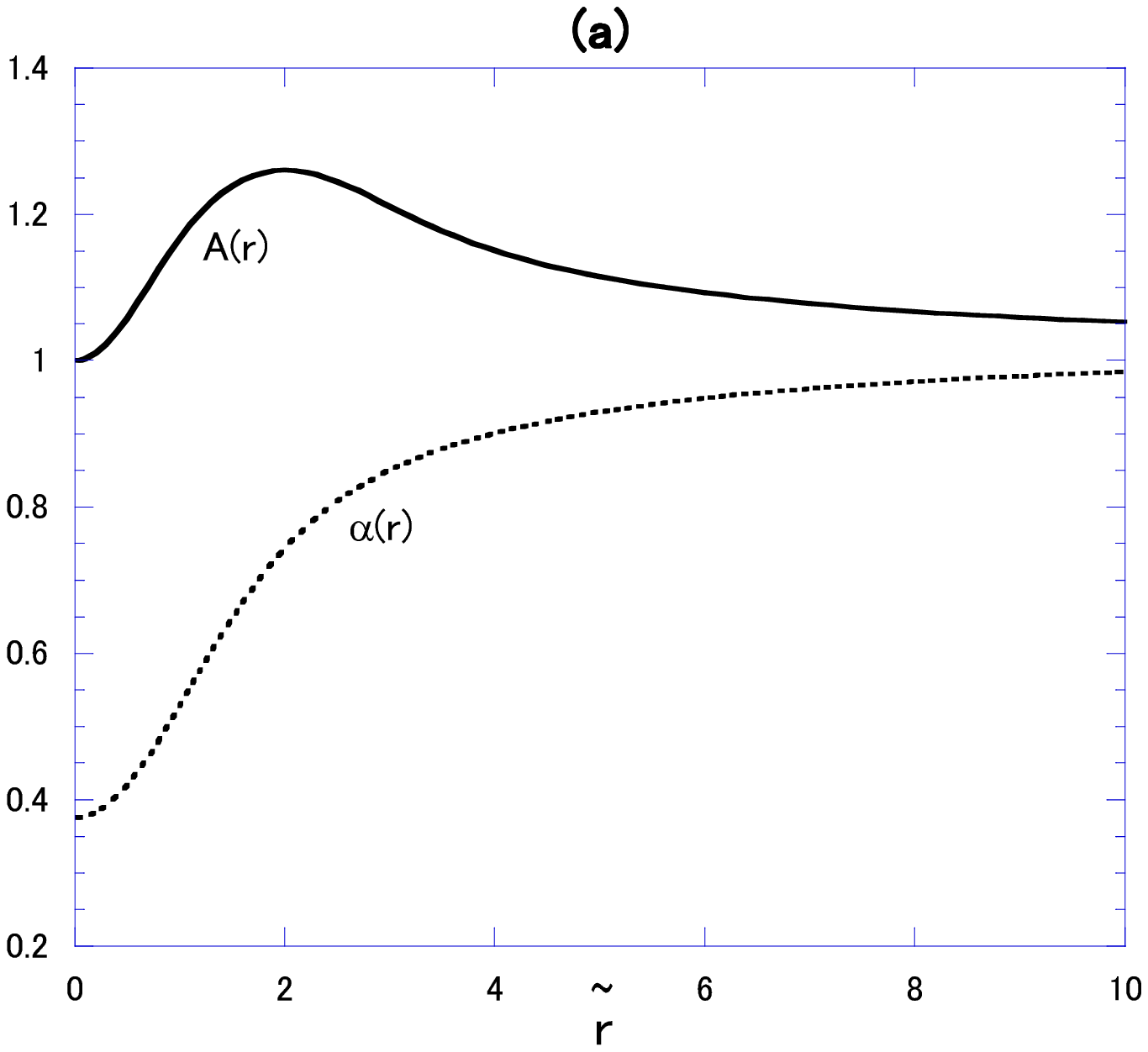,width=3in} \\
\psfig{file=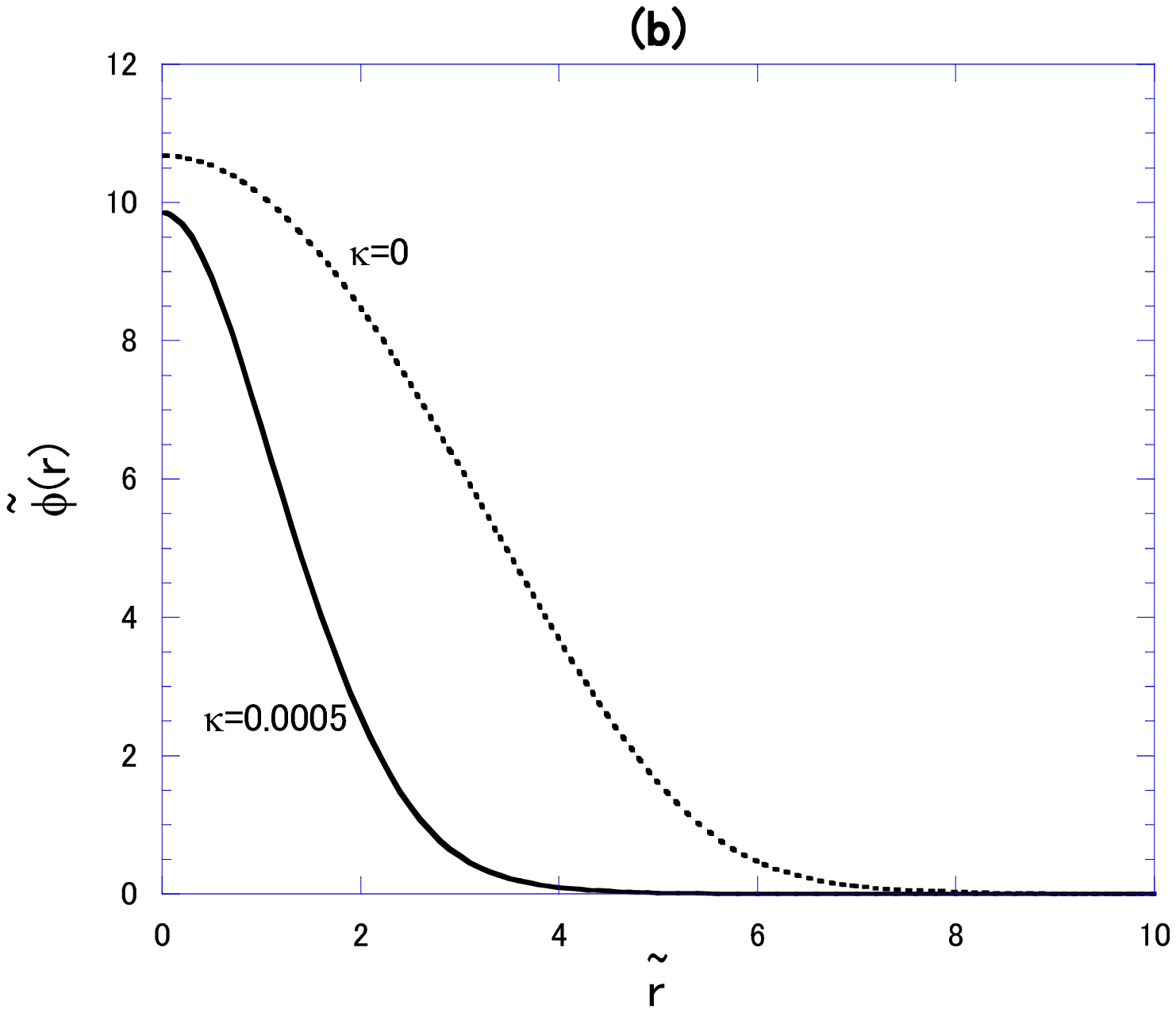,width=3in}
\caption{For a Q-ball for $\to^2 \simeq 0.35$, we show 
(a) behavior of the metric functions for a gravitating one for $\kappa =0.0005$ and 
(b) comparison of $\tp (r)$ with that for the flat case $\kappa =0$. 
\label{thin-behavior} }
\end{figure}
\begin{figure}[htbp]
\psfig{file=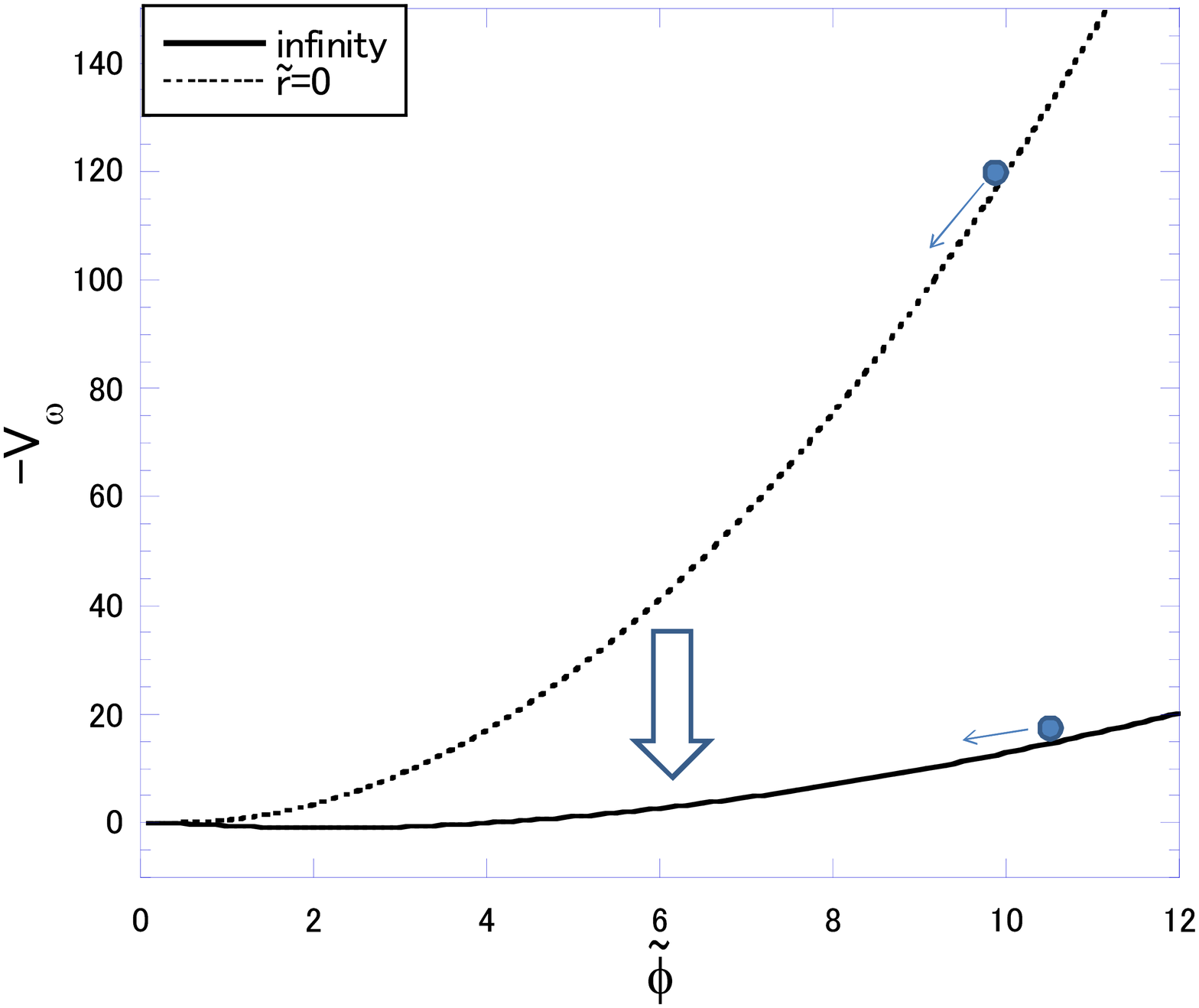,width=3in} 
\caption{$-V_{\omega}$ for a Q-ball $\to^2 \simeq 0.35$. 
For the flat case $\kappa =0$, $-V_{\omega}$ is fixed as shown by a 
solid line while $-V_{\omega}$ for a gravitating case $\kappa =0.0005$ changes 
from a dotted line to a solid line as ``time" $\tr$ goes. 
Since this change occurs ``quickly", their difference is small comparing with 
that might be expected. I.e., the scalar field 
$\tp \simeq 10.68$ ($\simeq 9.85$) at $\tr =0$ 
rolls down the potential and finally reaches $\tp =0$ at $\tr \ra \infty$ 
for the flat (gravitating) case. 
\label{Vw035} }
\end{figure}

Second, we show the result on a relatively thin Q-ball with $\to^2 \simeq 0.35$ in Fig.~\ref{thin-behavior}.
The metric functions $\alpha(\tr),~A(\tr)$ and the field amplitude $\tp(\tr)$ are shown in (a) and (b), respectively.
(a) tells us that, compared with the case $\to^2 \simeq 1.999$, the gravitational field is fairly stronger near the origin but approaches the flat spacetime faster as $\tr \ra \infty$. 
(b) indicates that the size of the Q-ball becomes smaller due to self-gravity.

We can also explain the reason for this result by using the potential $-V_{\omega}$, as shown in Fig.~\ref{Vw035}. 
For the flat case $\kappa =0$, $-V_{\omega}$ is fixed as shown by a 
solid line while $-V_{\omega}$ for a gravitating case $\kappa =0.0005$ changes 
from a dotted line to a solid line as ``time" $\tr$ goes. 
An essential difference from the former thick-wall case is that this change occurs ``quickly". 
As a result, the scalar field $\tp \simeq 10.68$ ($\simeq 9.85$) at $\tr =0$ 
rolls down the potential and finally reaches $\tp =0$ at $\tr \ra \infty$ 
for the flat (gravitating) case. Thus, $\tp$ for the gravitating case rolls down 
the potential faster than that for the flat case. 

Larger influence caused by weak gravity for the thick-wall case might seem paradoxical. 
To understand it, it is convenient to define the density $\rho$ and the radial pressure $p_{r}$ 
of the scalar field  in the fluid form as  
\beq\label{density}
\rho :={{\phi'}^2\over2A^2}+{\omega^2\phi^2\over2\alpha^2}+V_{\rm gauge},
\eeq
\beq\label{pressure}
p_{r}:={{\phi'}^2\over2A^2}+{\omega^2\phi^2\over2\alpha^2}-V_{\rm gauge}. 
\eeq
Then, let us remember hydrostatic equilibrium equations, which are another expression 
of Einstein equations, as in the usual star. 
\beq\label{mass}
\frac{dm(r)}{dr}=4\pi r^2 \rho ,
\eeq
\beq\label{pressure-gradient}
\frac{dp_{r}}{dr}=-\frac{Gm(r)\rho}{r^2}, 
\eeq
where $m(r)$ is the mass function of the Q-ball. 
We should notice that the pressure gradient must work as a repulsive 
force against the gravity to support the Q-ball while it works as an attractive force in the flat case. 
If we pay attention to the values of $\tr$ and $\tp$ in Figs.~\ref{thick-behavior} 
and \ref{thin-behavior} (b), we find that the absolute value of the pressure gradient 
for the thick-wall case is far smaller than that for the thin-wall case. 
Thus, weaker gravity does not necessarily mean a smaller influence to Q-balls and 
it is interesting to investigate its influences for various $\to^2$ 
which will be discussed below. 

\begin{figure}[htbp]
\psfig{file=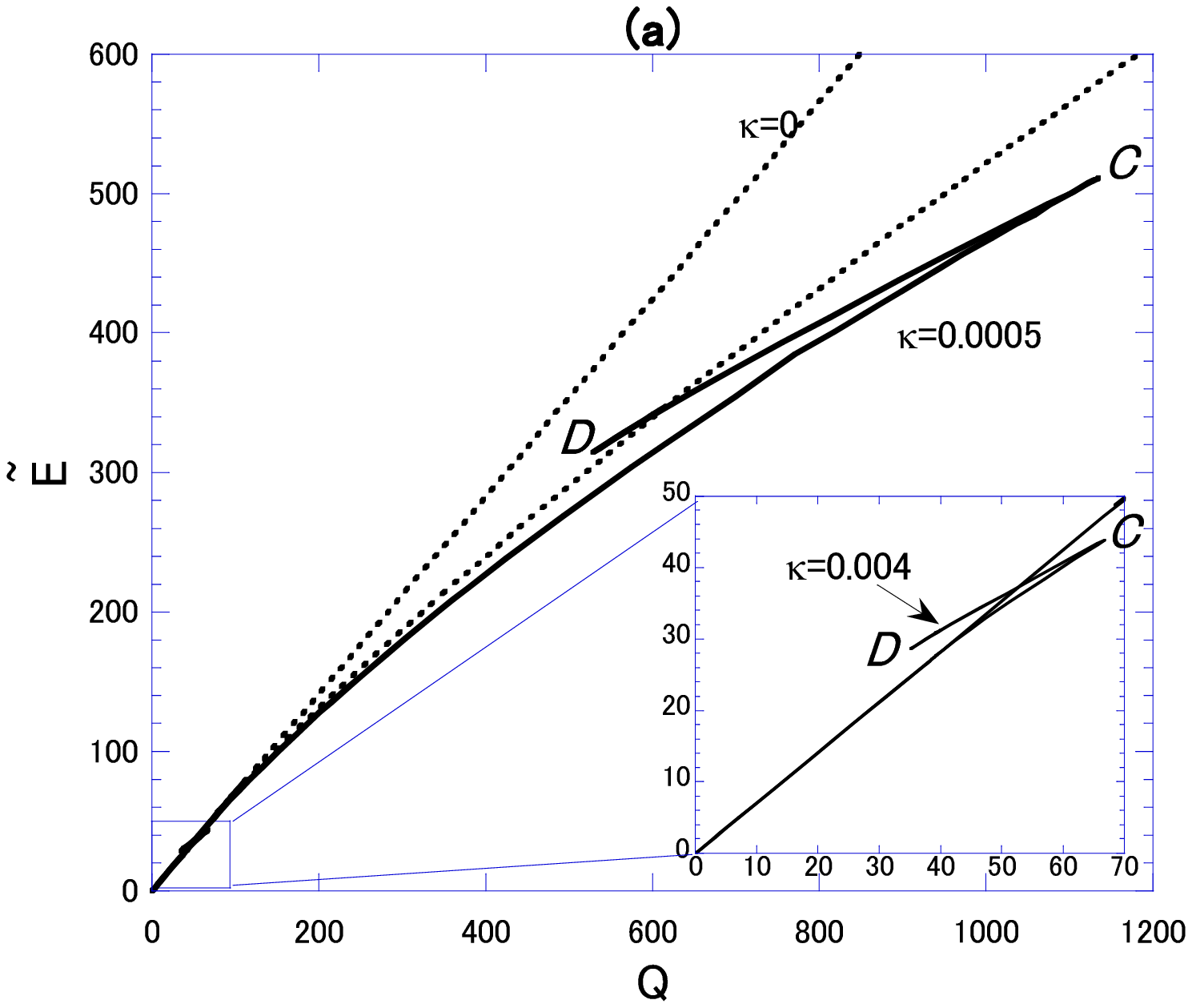,width=3in} \\
\psfig{file=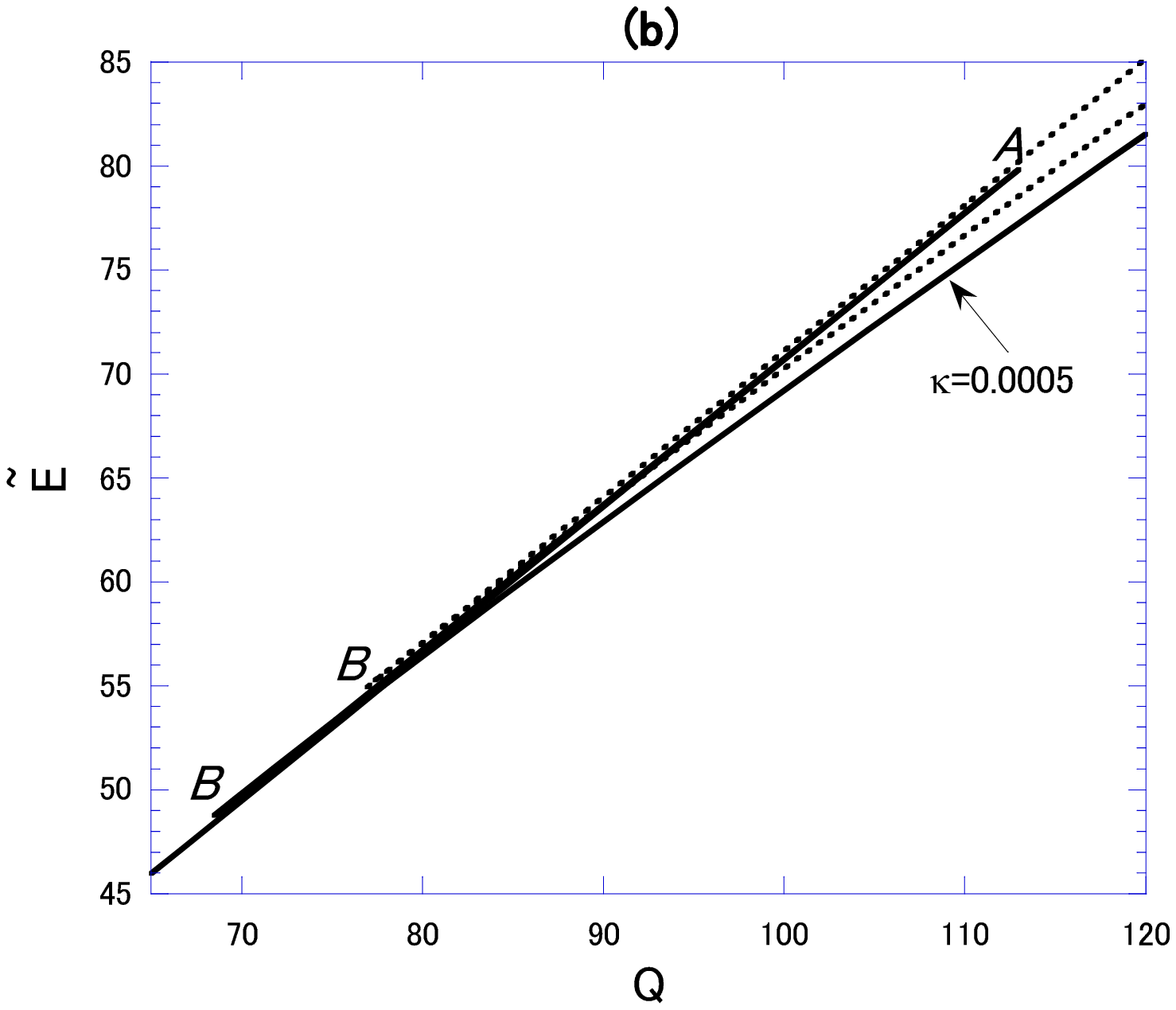,width=3in}
\caption{(a) $Q$-$\tE$ relation for $\kappa =0$, $0.0005$ and $0.004$ 
and (b) its magnification around $Q\sim 100$. 
\label{QE} }
\end{figure}
\begin{figure}[htbp]
\psfig{file=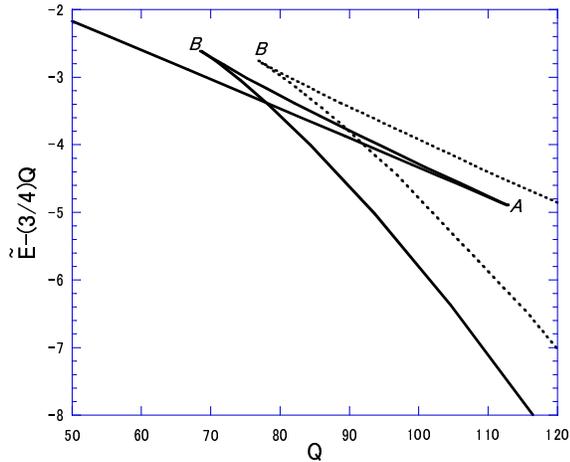,width=3in} 
\caption{$Q$-$(\tE -\frac{3}{4}Q)$ relation corresponding to Fig.~\ref{QE} (b). 
\label{E-34Q} }
\end{figure}
\begin{figure}[htbp]
\psfig{file=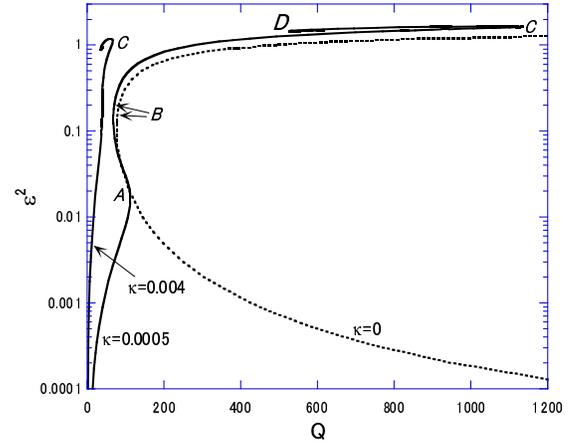,width=3in} 
\caption{$Q$-$\epsilon^2$ relation for $\kappa =0$, $0.0005$ and $0.004$. 
\label{Qepsilon} }
\end{figure}

As we discussed in our previous papers~\cite{TamakiSakai,TamakiSakai2,TamakiSakai3}, stability of 
Q-balls can be easily understood from the relation between $Q$ and 
the Hamiltonian energy $E$, which is defined by 
\beq\label{H}
E=\lim_{r\ra\infty}{r^2\alpha'\over2GA}={M_S\over2},
\eeq
where $M_S$ is the Schwarzschild mass. 
Here, stability means local stability, that is, stability against small perturbations. 
We also normalize $E$ as
\beq\label{Enormalize}
\tE := \frac{E}{m}. 
\eeq
We show $Q$-$\tE$ relation for $\kappa =0$, $0.0005$ and $0.004$ in Fig.~\ref{QE} (a) 
and its magnification around $Q\sim 100$ in (b). Since complicated structures are 
concentrated near the line $\tE =\frac{3}{4}Q$ in (b), we also show the corresponding 
$Q$-$(\tE -\frac{3}{4}Q)$ relation in Fig.~\ref{E-34Q}. We find that this relation is 
similar to that for the $V_{4}$ model (see, Fig.3 in \cite{TamakiSakai2}).
Therefore, the stability for the present model can be also understood in the same way as that for the $V_{4}$ model.

For the flat case represented by the dotted line, there are double values of $\tE$ for a given value of $Q$.
These two branches merge at the point $B$ ($Q\sim 77$) as shown in (b) or Fig.~\ref{E-34Q}. 
We can understand that the upper(lower) branch represents unstable (stable) solutions. 

Next, we discuss the case of $\kappa =0.0005$. 
Figures \ref{QE} and Fig.~\ref{E-34Q} tells us that the solution sequence $A$-$B$-$C$ is analogous to the sequence in the flat case.
We can therefore interpret the upper (lower) branch from $A$ to $B$ (from $B$ to $C$) 
represents unstable (stable) solutions. 
On the other hand, we can find the intrinsic differences 
from the flat case from $A$ to the origin and from $C$ to $D$ (and sequences of cusp structures). 
By energetic (or catastrophic) argument described in our previous paper~\cite{TamakiSakai2}, 
the branch from $A$ to the origin is regarded as stable. 
The point $C$ corresponds to the maximum of $Q$ ($Q_{\rm max}$). 
This suggests that a Q-ball with larger charge than $Q_{\rm max}$ cannot support itself 
due to the self-gravity. 
Both this interpretation and  energetic (or catastrophic) argument indicates that the 
solutions $C$ to $D$ (and sequences of cusp structures) are unstable. 

As for the case $\kappa =0.004$, $Q_{\rm max}$ becomes very small due to 
the large self-gravity and there is no fine structure like the sequence $A$-$B$ for $\kappa =0.0005$.
The smooth curve from the origin to the point $C$ represents stable solutions.
 
As we discussed in \cite{TamakiSakai,TamakiSakai2,TamakiSakai3}, $\to^2$ or $\epsilon^2$ is a {\it state variable}, while $Q$ and $\kappa$ are {\it control parameters}, in words of catastrophe theory.
Therefore it is instructive to depict the $Q$-$\epsilon^2$, too, in Fig.~\ref{Qepsilon}.
For the thick-wall solutions $\epsilon^2 \ll 1$ with gravity, 
as we explained using Fig.~\ref{Vw1999}, Q-ball charge is very small because $\tp(0)\ll 1$,
and there is no lower bound of $Q$,  contrary to the flat case. 
The solution near the point $C$ was also explained  using Fig.~\ref{Vw035}. 
In this case, the Q-ball charge becomes small due to the strong gravity. 
These phenomena occur for $|g^{rr}-1|\sim 1$ at its peak~\cite{TamakiSakai,TamakiSakai2}. 
For $\kappa =0.004$, the thick-wall regime and the strong gravity regime 
overlap each other. Therefore, the solution sequence is quite different from that for the flat case.

\begin{figure}[htbp]
\psfig{file=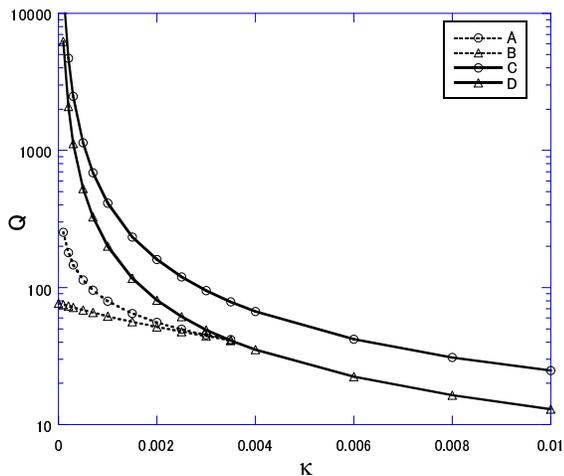,width=3in} 
\caption{Values of $Q$ for various $\kappa$ 
corresponding to the points $A$, $B$, $C$ and $D$ in Fig.~\ref{QE}. 
\label{kappa} }
\end{figure}

Because the extremal points $A,~B,~C$ and $D$ in Fig.~\ref{QE} indicate the main feature 
of the solutions sequences, we also depict how the values of $Q$ at these points vary with $\kappa$ in Fig.~\ref{kappa}.
The values of $Q$ at $C$, which corresponds to $Q_{\rm max}$, become smaller as $\kappa$ becomes larger; 
this is a common feature with $V_{3}$ and $V_{4}$ models~\cite{TamakiSakai,TamakiSakai2,multamaki}. 
The values of $Q$ at the points $A$ and $B$ merge at $\kappa \simeq 0.0035$ and 
the small structure disappears for $\kappa >0.0035$, as already seen for $\kappa =0.004$ in Fig.~\ref{QE}.  
These continuous change as $\kappa$ is easily understood.

This figure, however, reveals a nontrivial result.
The local maximum $A$ does not disappear for $\kappa\ra+0$ while it is nonexistent for $\kappa=0$.
This means that there is no lower bound of $Q$ for $\kappa\ra+0$ while there is a minimum of $Q$ for $\kappa=0$.
This result is against our naive idea that gravitational effects should vanish in the limit of $\kappa\ra+0$. 
Therefore, It should be argued carefully whether or not there remain thick-wall solutions with $Q\ra 0$ in the weak gravity limit $\kappa \neq 0$. This is the subject of the next section. 

\section{Thick-wall solutions for $\kappa \neq 0$}
We consider thick-wall solutions ($\epsilon^2\ll1$) with weak gravity 
by expressing the metric functions as 
\beq\label{weak}
\alpha^{2}=1+h(r), ~~A^2=1+f(r),~~
(|h|\ll1, ~~ |f|\ll1),
\eeq
and we shall take up to first order in $h$ and $f$ hereafter. 
As we discussed for the flat spacetime in Sec.\ IIB, we evaluate $\phi_{0}:=\tp (0)$ as a solution of 
$V_{\omega}=0$ with (\ref{weak}), 
\beq\label{phi0}
\epsilon^2 +\to^2 h(0)-\phi_{0}^2\simeq 0, 
\eeq
where we have neglected higher order terms $O(\phi_{0}^{5})$.

Let us consider the limit $\epsilon^2\ra0$. 
For the flat case $\kappa =0$, $\phi_{0}\simeq \epsilon$ since $h(0)$ can be taken to 
be zero identically. However, for any small $\kappa \neq 0$, 
it is not evident whether or not $h(0)$ can be negligible and we should compare 
the order of $h(0)$ with that of $\epsilon^2$. 
For this purpose, we should also estimate it by using the Einstein equations.
To do this, we assume a top-hat configuration,
\beq\label{h-estimate0}
\tp (\tr )\sim \phi_{0}\ll1
~~ {\rm for }~~ \tr <\frac{C}{\epsilon},\ \ C={\rm const. }
\eeq
For the flat case, $\phi_{0}=\epsilon$, as in Eq.(\ref{phi0-flat2}).

\begin{figure}[htbp]
\psfig{file=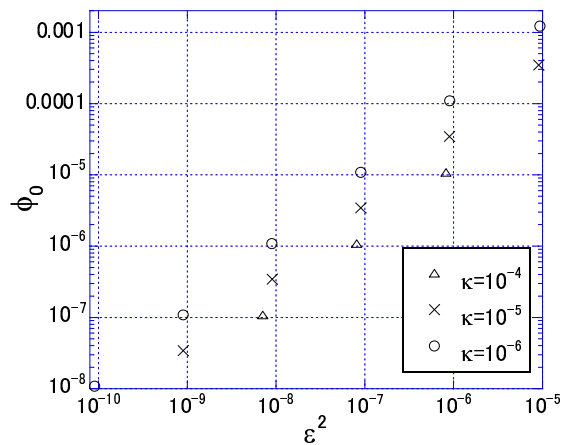,width=3in} 
\caption{$\epsilon^2$-$\phi_{0}$ relation for various $\kappa$. 
\label{phi0num} }
\end{figure}
From the Einstein equations, we find
\beq\label{h-estimate}
-G^t_t+G^i_i:=\left({\tr^2 \alpha '\over A}\right)'
=8\pi\kappa \tr^2 A\alpha \left(\frac{\to^2 \tp^2}{\alpha^2} -V\right)\ ,
\eeq
where $i$ denotes spatial components. 
If we take the weak field approximation (\ref{weak}) and the thick-wall approximation $\epsilon^2\ll1$, we obtain
\beq\label{h-estimate2}
(\tr^2 h ')'\simeq 16\pi\kappa \tr^2 \tp^2 \ .
\eeq
With the central boundary condition $h'(0)=0$ and the approximation (\ref{h-estimate0}), 
we can integrate (\ref{h-estimate2}) as 
\beq\label{h-estimate3-1}
h'\simeq \frac{16}{3}\pi\kappa  \phi_{0}^2 \tr,
~~ {\rm for }~~ \tr <{C\over\epsilon}.
\eeq
With the outer boundary condition $h(\frac{C}{\epsilon})\simeq h(\infty)=0$, 
we can integrate (\ref{h-estimate3-1}) as 
\beq\label{h-estimate3}
h(0)\simeq -\frac83\pi\kappa \phi_{0}^2\frac{C^2}{\epsilon^2}.
\eeq
From (\ref{phi0}) and (\ref{h-estimate3}), we obtain 
\beq\label{phi0-2}
\phi_{0}^{2}=\frac{3\epsilon^4}{8\pi\kappa C^2 +3\epsilon^2}. 
\eeq

This formula clearly shows how $\phi_0$ converges as $\epsilon\ra0$ and $\kappa\ra0$;
it depends on their convergent rates.
If $\epsilon^2\gg\kappa C^2$, we have $\phi_{0}\simeq \epsilon$, as in the flat case.
On the other hand,  if $\epsilon^2\ll\kappa C^2$, we have 
\beq
\phi_{0}\simeq \frac{\epsilon^2}{2C}\sqrt{\frac{3}{2\pi\kappa}}.  \label{phi0-3}
\eeq
In the real situation, $\kappa$ is very small but a nonzero constant, while $\epsilon$ is variable and determined by initial conditions.
Therefore, if we discuss the thick-wall limit in weak gravity, the latter result (\ref{phi0-3}) applies.
In this case, because
\beq
Q\propto \tp^{2}\tr^3\sim\phi_{0}^{2}{1\over\epsilon^3}\sim\epsilon,
\eeq
there is no lower bound of $Q$ as expected. 
Moreover, the point $A$ in Fig.~\ref{Qepsilon} can be interpreted as the point when
$\epsilon^2\sim\kappa C^2$.
To confirm the above argument, we show the numerical relation $\epsilon^{2}$-$\phi_{0}$ in Fig.~\ref{phi0num}. 
We find that (\ref{phi0-3}) holds true and $C\sim O(10)$. 

\begin{figure}[htbp]
\psfig{file=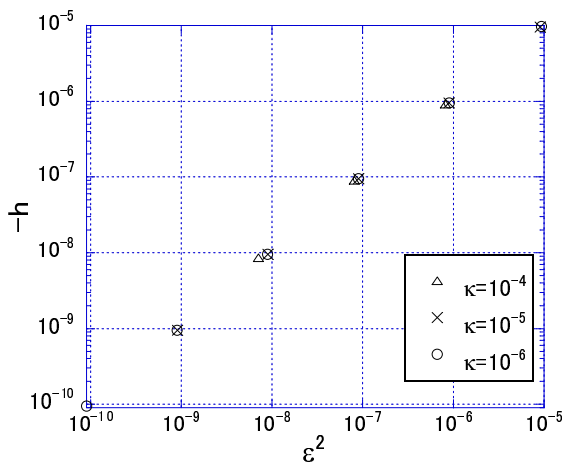,width=3in} 
\caption{$\epsilon^2$-$(-h(0))$ relation for various $\kappa$. 
\label{h0num} }
\end{figure}

To check the consistency with the assumption of weak gravity, we also estimate $h(0)$ by
substituting (\ref{phi0-3}) into (\ref{phi0}),
\beq\label{h-estimate4}
h(0)\simeq -\epsilon^{2}\ .
\eeq
Because of the thick-wall assumption $\epsilon\ll 1$, it is consistent with the first assumption
$|h|\ll1$, and confirmed by the  numerical relation $\epsilon^{2}$-$(-h(0))$ in Fig.~\ref{h0num}. 
This result tells us that $h(0)$ cannot be negligible in Eq.(\ref{phi0}) independent of $\kappa\neq 0$. 
Thus, we should distinguish between $\kappa =0$ (i.e., $h(0)=0$) and 
any small value of $\kappa \neq 0$ clearly. 
In summary, if $\epsilon^2\ll\kappa C^2$ and $\epsilon\ll 1$ are satisfied, 
we conclude that effects of self-gravity cannot be ignored no matter how weak the gravity is. 

We want to examine what types of potentials we can apply the above results. 
To obtain (\ref{phi0-2}), we have assumed that the second leading order is quartic, $\tp^4$.
However, to obtain (\ref{phi0-3}), which is the result for $\epsilon^2\ll\kappa C^2$, we have only assumed that the leading order in $V_{\omega}$ (or that in the Maclaurin series) is quadratic, $\tp^2$.
Therefore, the above results hold true not only for the model (\ref{gauge}) but also general models 
in which a positive mass term is a leading order one. 
 
\section{Conclusion and discussion}
We have investigated how gravity affects Q-balls by exemplifying the case 
with the AD potential $V(\phi):=m^4 \ln (1+\frac{\phi^2}{m^2})$. 
Surprisingly, stable Q-balls with arbitrarily small charge exist, 
no matter how weak gravity is, contrary to the case of flat spacetime. 
The result for the gravitating case is a universal property which has been known 
to hold for various potentials~\cite{boson-review,TamakiSakai,TamakiSakai2}. 
We have also showed that this feature of gravitating Q-balls holds true for general models as long as 
the leading order term of the potential (or that in the Maclaurin series) is a positive mass term. 
Therefore, this result does not change if we include nonrenormalization terms, which we have ignored in 
the gauge-mediated AD potential.
Our results suggest that gravity may play an important role in the Q-ball formation process.

\acknowledgements
We would like to thank Kei-ichi Maeda for continuous encouragement. 
The numerical calculations were carried out on SX8 at  YITP in Kyoto University. 
This work was supported by MEXT Grant-in-Aid for Scientific Research on Innovative Areas No.\ 22111502.



\begin{thebibliography}{99}
\bibitem{Col85}
S. Coleman, Nucl. Phys. {\bf B262}, 263 (1985).
\bibitem{LP92}For a review of nontopological solitons in flat spacetime, see, 
T. Lee and Y. Pang, Phys. Rep. {\bf 221}, 251 (1992). 
\bibitem{AD}I. Affleck and M. Dine, Nucl. Phys. B {\bf249} 361 (1985).
\bibitem{SUSY}
A. Kusenko, Phys. Lett. B {\bf405}, 108 (1997) 108; Nucl. Phys. B (Proc. Suppl.) 62A-C, 248 (1998); 
K. Enqvist and J. McDonald, Phys. Lett. B {\bf 425}, 309 (1998); Nucl. Phys. B {\bf 538}, 321 (1999); 
S. Kasuya and M. Kawasaki, Phys. Rev. D {\bf62}, 023512 (2000).
\bibitem{SUSY-DM}
A. Kusenko and M. Shaposhnikov, Phys. Lett. B {\bf418}, 46 (1998);
K. Enqvist and A. Mazumdar, Phys. Rep. {\bf 380}, 99 (2003); 
I. M. Shoemaker and A. Kusenko, Phys. Rev. D {\bf80}, 075021 (2009).
\bibitem{stability}
A. Kusenko, Phys. Lett. B {\bf404}, 285 (1997); {\bf406}, 26 (1997);
F. V. Kusmartsev, Phys. Rep. {\bf 183}, 1 (1989). 
T. Multamaki and I. Vilja, Nucl. Phys. B {\bf 574}, 130 (2000);
M. Axenides, S. Komineas, L. Perivolaropoulos and M. Floratos, Phys. Rev. D {\bf 61}, 085006 (2000).
\bibitem{PCS01}F. Paccetti Correia and M. G. Schmidt, Eur. Phys. J. {\bf C21}, 181 (2001).
\bibitem{SakaiSasaki}N. Sakai and M. Sasaki, Progress of Theoretical Physics, {\bf 119}, 929 (2008).
\bibitem{Copeland}
M. Gleiser and J. Thorarinson, Phys. Rev. D {\bf 73}, 065008 (2006); 
M. I. Tsumagari, E. J. Copeland, and P. M. Saffin, Phys. Rev. D {\bf 78}, 065021 (2008); 
E. J. Copeland and M. I. Tsumagari, Phys. Rev. D {\bf 80}, 025016 (2009). 
\bibitem{boson-review}For a review of boson stars, see, 
P. Jetzer, Phys. Rep. {\bf 220}, 163 (1992). 
F. E. Schunck and E. W. Mielke, Class. Quantum Grav. {\bf 20}, R301 (2003).
\bibitem{TamakiSakai}
T. Tamaki and N. Sakai, Phys. Rev. D {\bf81}, 124041 (2010). 
\bibitem{TamakiSakai2}
T. Tamaki and N. Sakai, Phys. Rev. D {\bf83}, 044027 (2011). 
\bibitem{TamakiSakai3}
T. Tamaki and N. Sakai, Phys. Rev. D {\bf 83}, 084046 (2011). 
\bibitem{multamaki}
T. Multamaki and I. Vilja, Phys. Lett. B {\bf 542}, 137 (2002). 

\end{thebibliography}
\end{document}